\documentstyle[twocolumn,prl,aps]{revtex}
\input{psfig.sty}

\begin{document}
\draft

\author{A. R. Denton$^1$ and H. L\"owen$^{1,2}$}
\address
{$^1$~Institut f\"ur Festk\"orperforschung, Forschungszentrum J\"ulich,
D-52425 J\"ulich, Germany}
\address
{$^2$~Institut f\"ur Theoretische Physik II, Universit\"at D\"usseldorf,
D-40225 D\"usseldorf, Germany}

\title{Stability of Colloidal Quasicrystals}

\date{\today}
\maketitle

\begin{abstract}
Freezing of charge-stabilized colloidal suspensions and relative 
stabilities of crystals and quasicrystals are studied using 
thermodynamic perturbation theory.  Macroion interactions 
are modelled by effective pair potentials combining electrostatic 
repulsion with polymer-depletion or van der Waals attraction.  
Comparing free energies -- counterion terms included -- for 
elementary crystals and rational approximants to icosahedral quasicrystals, 
parameters are identified for which {\it one-component} quasicrystals 
are stabilized by a compromise between packing entropy and cohesive energy.
\end{abstract}




Suspensions of mesoscopic-sized ($1-1000$ nm) colloidal particles dispersed
in a fluid medium are of both practical and fundamental interest
~\cite{Pusey}.
Beyond traditional relevance to the chemical, food, and pharmaceutical
industries, the remarkable mechanical, thermal, and optical properties of 
these typically soft materials raise intriguing prospects for novel 
applications, such as optical switching devices~\cite{Asher}.
Recent scientific interest has been driven largely by advances in
sample preparation, scattering, and imaging techniques~\cite{Mainz}.
As well-characterized models of atomic systems, colloidal suspensions offer 
valuable insight into the basic link between microscopic interparticle 
interactions and macroscopic phase behavior in condensed matter.
More so than in atomic systems, colloidal interactions are eminently 
tunable through experimental control of particle size, charge, and 
surface chemistry, as well as properties of the suspending medium, 
such as polarizability and salt concentration.  
A correspondingly rich variety of thermodynamic phases has been observed, 
including the vapor, liquid, crystal, and glass phases common to atomic 
systems.  A notable exception is the quasicrystal phase, 
which to date is known to exist only in certain binary and ternary 
metallic alloys~\cite{QC1}.
The purpose of this Letter is to demonstrate, using general statistical 
mechanical methods, the possibility of thermodynamically stable 
{\it one-component} colloidal quasicrystals.
If experimentally observed, such systems would represent a fundamentally
new type of colloidal order and permit the first real-space imaging 
of quasicrystals. 

We focus specifically on suspensions of charged spherical colloidal 
macroions and oppositely charged counterions dispersed in 
a deionized (salt-free) solvent.
The macroions are modelled as hard spheres of diameter $\sigma$ and 
uniform surface charge $Ze$, the counterions as point particles of 
elementary charge $e$, and the solvent as a continuous medium 
of dielectric constant $\epsilon$.
Charge neutrality relates the average macroion and counterion number 
densities via $Z\rho_m=\rho_c$.
Electrostatic interactions between charged macroions in suspension 
are commonly modelled by an effective hard-sphere/screened-Coulomb 
pair potential: 
\begin{eqnarray}
\phi_{el}(r)~&=&~\frac{Z^2e^2}{\epsilon}\Bigl[\frac{\exp(\kappa\sigma/2)}
{1+\kappa\sigma/2}\Bigr]^2~\frac{\exp(-\kappa r)}{r},\qquad r>\sigma
\nonumber\\
 ~&=&~\infty,\qquad\qquad\qquad\qquad\qquad\qquad\qquad r<\sigma,
\label{DLVO}
\end{eqnarray}
where $r$ is the separation between macroion centers and
$\kappa=\sqrt{4\pi Ze^2\rho_m/\epsilon k_BT}$
is the Debye screening constant.
Equation (\ref{DLVO}) was first derived in the dilute regime
by Derjaguin, Landau, Verwey, and Overbeek~\cite{DLVO}. 
The screened-Coulomb form has since been shown to be accurate 
also for concentrated suspensions, with $Z$ replaced by an effective 
charge~\cite{Alexander}.
At finite macroion volume fractions $\eta=(\pi/6)\rho_m\sigma^3$, the 
volume available to the counterions is smaller than the total volume $V$
by a factor $(1-\eta)$.
As a result, the effective counterion density is increased by a factor 
$1/(1-\eta)$, {\it enhancing} screening~\cite{Russel} according to
$\tilde\kappa=\sqrt{4\pi Ze^2\rho_m/\epsilon k_BT(1-\eta)}$.
Henceforth, we assume an electrostatic potential of the form of
Eq.~(\ref{DLVO}), but with this modified screening constant.

Aside from electrostatic repulsions, 
interactions between macroions also can include an attractive component.  
Attractions may arise, for example, in the presence of free (nonadsorbing) 
polymer via a polymer-depletion mechanism~\cite{AO-Vrij}.
For macroion surface-surface separations smaller than the characteristic
polymer coil diameter, depletion of polymer from the intervening space 
creates an osmotic pressure imbalance drawing the macroions together.  
The attractive energy is directly proportional to the polymer osmotic 
pressure $\Pi_p$ and the mutual overlap of excluded volume shells. 
For coils of sufficiently small radius of gyration $R_g$ relative
to $\sigma$, the effect is described by a pair potential 
that is nonzero only in the range 
$\sigma < r < \sigma + 2R_g$, where it has the form~\cite{AO-Vrij}
\begin{equation}
\phi_{dep}(x)~=~-c\Bigl[1-\frac{3x}{2(1+\xi)}
+\frac{1}{2}\Bigl(\frac{x}{1+\xi}\Bigr)^3\Bigr],
\label{dep}
\end{equation}
with $x=r/\sigma$, $\xi=2R_g/\sigma$, and $c=(\pi/6)\Pi_p\sigma^3(1+\xi)^3$.
For size ratios $\xi>0.16$, triplet overlaps entail three-body interactions. 
However, for $\xi \simeq 0.25$ (our case below) these are small enough 
to be ignored~\cite{Gast1}.
Attractions also may arise from fluctuating 
dipole-dipole forces, as described by the London-van der Waals 
pair potential~\cite{DLVO}
\begin{equation}
\phi_{vdW}(x)~=~-\frac{A}{6}\Bigl[\frac{1}{2x^2}+\frac{1}{2(x^2-1)}+
\ln(1-\frac{1}{x^2})\Bigr],
\label{vdW}
\end{equation}
where the Hamaker constant $A$ is related to the macroion and solvent 
polarizabilities (refractive indices).
The negative divergence at contact ($r=\sigma$) can cause
irreversible coagulation unless the macroions are stabilized, 
{\it e.g.}, sterically by adsorbing or grafting a layer of polymer 
to their surfaces.  For simplicity, we model this by a cut-off 
at a distance $r=R_c$, roughly $5-10$ \% larger than $\sigma$.

Figure~1 depicts effective macroion-macroion pair potentials that combine 
electrostatic repulsion with each of the two types of attraction 
discussed above. 
The length scales of the interactions are set by the macroion core diameter,  
the polymer coil radius (or van der Waals cut-off distance), and the
Bjerrum length $\lambda_B\equiv e^2/\epsilon k_BT$ (at temperature $T$).
These pair potentials are now taken as input to a statistical mechanical
theory for the Helmholtz free energy.
The macroion contribution to the free energy is determined by means of 
thermodynamic perturbation theory~\cite{HM}.  
(An earlier such study by Gast {\it et al.}~\cite{Gast2} considered 
freezing only into an FCC crystal.)
The theory proceeds by splitting the pair potential 
into short-range reference and longer-range perturbation potentials.
The natural separation here is into a hard-sphere (HS) reference potential 
$\phi_{HS}(r)$ and a perturbation potential $\phi_p(r)$ that combines
an attractive well with a repulsive barrier and long-range tail.
To first order in $\phi_p(r)$, the macroion free energy separates 
correspondingly into reference and perturbation terms:
\begin{eqnarray}
F_m[\rho_m({\bf r})]~=~F_{HS}[\rho_m({\bf r})]
\nonumber\\
~+~2\pi\rho_mN_m\int_0^{\infty}{\rm d}r'~r'^2~
g_{HS}(r';[\rho_m({\bf r})])~\phi_p(r'),
\label{pert}
\end{eqnarray}
$N_m=\rho_mV$ being the macroion number, and $F_{HS}[\rho_m({\bf r})]$ and 
$g_{HS}(r;[\rho_m({\bf r})])$ the free energy and radial distribution 
function (RDF), respectively, of the HS reference system, both dependent on 
the {\it equilibrium} macroion density distribution $\rho_m({\bf r})$. 
Accuracy of first-order perturbation theory is assured if fluctuations 
in the total perturbation energy remain small relative to $k_BT$~\cite{HM}, 
a condition generally well satisfied here, where variations in $\phi(r)$
are small for distances over which the RDF varies appreciably.

For the fluid phase, with spatially constant macroion density, 
the macroion free energy is calculated via the uniform limit 
[$\rho_m({\bf r}) \to \rho_m$] of Eq.~(\ref{pert}), using the 
essentially exact Carnahan-Starling and Verlet-Weis forms~\cite{HM} 
for $F_{HS}(\rho_m)$ and $g_{HS}(r;\rho_m)$, respectively.
For the solid phase, classical density-functional methods~\cite{DFT}
are applied.
The reference free energy is obtained by minimizing a {\it functional} 
${\cal F}_{HS}[\rho_m({\bf r})]$ with respect to the 
macroion density distribution $\rho_m({\bf r})$, or in practice 
the width parameter $\alpha$ in the Gaussian parametrization
\begin{equation}
\rho_m({\bf r})~=~\left(\frac{\alpha}{\pi}\right)^{3/2}~\sum_{\bf R}
\exp(-\alpha|{\bf r}-{\bf R}|^2),
\label{Gauss}
\end{equation}
the sum running over the lattice sites ${\bf R}$ of a 
specified solid structure.
For non-overlapping Gaussians, 
the configurational entropy (ideal-gas) part of the functional 
is given exactly by ${\cal F}_{id}/N_mk_BT=(3/2)\ln(\alpha\sigma^2/\pi)-5/2$.
The excess part is determined by means of the 
modified weighted-density approximation (MWDA)~\cite{MWDA}, 
which reasonably describes HS solids.  
This maps the excess part of ${\cal F}_{HS}[\rho_m({\bf r})]/N_m$ onto its 
fluid counterpart, evaluated at an effective density, via
${\cal F}_{ex}^{MWDA}[\rho_m({\bf r})]/N_m~=~f_{ex}(\hat\rho)$, where
$f_{ex}$ is the fluid excess free energy per particle and $\hat\rho$
is a weighted average of $\rho_m({\bf r})$ that incorporates exact
two-particle correlations and a subset of higher-order correlations.
The reference RDF -- an angular average of the two-particle density --
is calculated using an approach recently proposed by
Rasc\'on {\it et al}.~\cite{Rascon}.
This corrects the first peak for nearest-neighbor correlations 
-- by fixing the contact value (via the virial theorem), 
coordination number, and first moment -- and treats 
higher-order peaks in a mean-field fashion.
The approximation compares excellently with simulation data for HS crystals 
and has been successfully applied to Lennard-Jones and square-well 
solids~\cite{Rascon}.

The counterion contribution to the free energy is~\cite{vRH}
\begin{equation}
\frac{F_c(\rho_m)}{N_m}~=~Zk_BT\ln \Bigl(\frac{\rho_m}{1-\eta}\Bigr)~-~
\frac{Z^2e^2}{2\epsilon}\frac{\tilde\kappa}{1+\tilde\kappa\sigma/2},
\label{vol}
\end{equation}
comprising the entropy of the counterions and the interaction of a 
macroion with its own counterion screening cloud.
Although independent of macroion structure, the counterion free energy
is manifestly state-dependent and, under conditions of high charge and 
low salt, can profoundly influence phase behavior.

Applying the above procedure, we have calculated total free energies for
a variety of elementary crystals and model quasicrystals.  
The crystalline structures examined include face-centered and body-centered
cubic (FCC and BCC), tetragonal, orthorhombic, and rhombohedral Bravais 
lattices, and the hexagonal close-packed (HCP) structure.
The model quasicrystalline structures are rational approximants (RA)
to icosahedral quasicrystals generated by projecting a six-dimensional 
hypercubic lattice onto the three-dimensional physical space and 
approximating, in the perpendicular space, the golden mean $\tau$ 
by a rational number $\tau_n=F_{n+1}/F_n$, where $F_n$ is a term 
in the Fibonacci sequence.
This procedure, together with a one-component occupation of the lattice 
sites that optimizes packing efficiency~\cite{RA}, yields periodic lattices 
with large unit cells and local order closely approximating that of 
aperiodic quasicrystals.
The first four RAs, denoted 1/1, 2/1, 3/2, and 5/3, have unit cells
containing, respectively, $20$, $108$, $452$, and $1904$ atoms
and maximum HS volume fractions of 
$0.5020$, $0.6400$, $0.6323$, and $0.6287$ ({\it cf}.~$0.6288$ 
in the quasiperiodic limit $n \to \infty$ 
and $0.7405$ for FCC and HCP crystals).  
The RDFs illustrated in Fig.~2 show that nearest-neighbor distances 
for the RAs are about $5$ \% {\it shorter} than for the crystals.  
For computational ease we focus on the 2/1 structure, but check 
to ensure qualitatively similar results for the higher-order approximants.

From previous work~\cite{DH}, packing efficiency is known to dominate
the stability of HS solids, the RAs being always metastable 
(higher in free energy) relative to the crystals.  
On the other hand, because of their shorter nearest-neighbor distances, 
the RAs can gain energetically from short-range attractive interactions.
In fact, for potential wells of appropriate depth (a few $k_BT$) 
and width ($~10-30$ \% of $\sigma$), their free energies can be 
lower than those of simple crystals. 
As shown above, such attractions can be produced in colloidal systems by 
adding polymer of appropriate size and concentration [Eq.~(\ref{dep})] 
or by tuning refractive indices [Eq.~(\ref{vdW})]. 

Assuming a room-temperature, salt-free, aqueous solvent ($\lambda_B=0.72$ nm),  
we have surveyed macroion parameters, $\sigma$ and $Z$, and parameters
defining attractive interactions -- $\xi$ and $\Pi_p$ for index-matched 
colloid-polymer mixtures, $R_c$ and $A$ for non-index-matched suspensions --
computed free energies for each solid structure, and identified 
parameter combinations tending to favor quasicrystal stability.
For illustration, Fig.~3 displays the free energies of a non-index-matched
suspension characterized by 
$\sigma=50$ nm, $Z=150 e$, $R_c/\sigma=1.058$, and $A/k_BT=25$.
Hamaker constants of this magnitude, though high for 
polymeric colloids, are typical for metallic colloids, such as Ag, Au, and Cu. 
Furthermore, charges of this magnitude correspond to counterion concentrations
of order $10^{-3}$ mol/l, far exceeding the background concentration of
$10^{-7}$ mol/l for H$^+$ and OH$^-$ ions from the dissociation of water.
Relative stabilities of the fluid and competing solid structures
are assessed by constructing Maxwell common tangents, ensuring 
equality of chemical potentials and pressures in coexisting phases.
Remarkably, over a range of densities intermediate between dense fluid 
and close-packed crystals, the quasicrystalline structures are 
predicted to be thermodynamically stable.
The stability can be traced to a competition between two factors, 
namely packing efficiency, favoring the FCC/HCP crystal, 
and nearest-neighbor coordination, favoring quasicrystalline order.  
This is best seen by comparing Figs.~1 and 2 and observing that
the attractive well of $\phi(r)$ is more commensurate with the 
first peak of $g_{HS}(r)$ for the 2/1 RA than for the FCC crystal.  
Stable quasicrystals thus emerge
as a compromise between packing entropy 
and cohesive energy in salt-free colloidal suspensions with a 
concentrated counterion background.

Applying the same approach to colloid-polymer mixtures, quasicrystalline 
stability is again predicted for sufficiently high polymer osmotic pressure. 
Repeating the coexistence analysis over a range of osmotic pressures, 
we have mapped out phase diagrams in the $\Pi_p$-$\rho_m$ plane.
A typical example is presented in Fig.~4. 
At lower osmotic pressures, where the interactions are purely repulsive, 
the fluid freezes into a close-packed crystal, FCC and HCP being 
essentially degenerate.
Beyond a threshold osmotic pressure, however, freezing occurs into the 
quasicrystal, which remains stable over a significant range of densities.  
At the threshold pressure (here $\Pi_p\sigma^3/k_BT\simeq 220$), which we
estimate to be near cross-over between dilute and semi-dilute polymer regimes, 
the fluid, crystal, and quasicrystal coexist at a triple point. 
At higher densities, packing efficiency prevails and the 
quasicrystal makes a structural transition to the FCC/HCP crystal.
The small relative density differences justify our assumption 
of equal polymer concentrations in coexisting phases.
Finally, we emphasize that the predicted region of quasicrystal stability 
is robust with respect to variation of parameters, 
widening and shifting to higher $\Pi_p$ with increasing $\sigma$ and $Z$, 
and that its existence depends not on specific details, but only 
the qualitative form, of the interactions.

Summarizing, using thermodynamic perturbation theory, we have investigated 
freezing transitions and assessed the relative stabilities of crystalline 
and quasicrystalline solids in charge-stabilized colloidal suspensions. 
For effective macroion interactions combining electrostatic repulsion 
with either polymer-depletion or van der Waals attraction, system parameters
and thermodynamic states have been identified for which one-component
icosahedral quasicrystals are predicted to be thermodynamically stable 
over a significant range of densities. 
Such systems and interactions should be experimentally achievable, 
raising prospects for observation of stable colloidal quasicrystals.

We thank Prof.~J.~Hafner and Dr.~A.M.~Denton for helpful discussions,  
and Drs.~M.~Kraj\^c\'i and M.~Windisch for supplying the rational 
approximant structures.




\newpage
\onecolumn




\unitlength1mm

\begin{figure}
\begin{center}
\begin{picture}(140,100)
\put(5,0){\psfig{figure=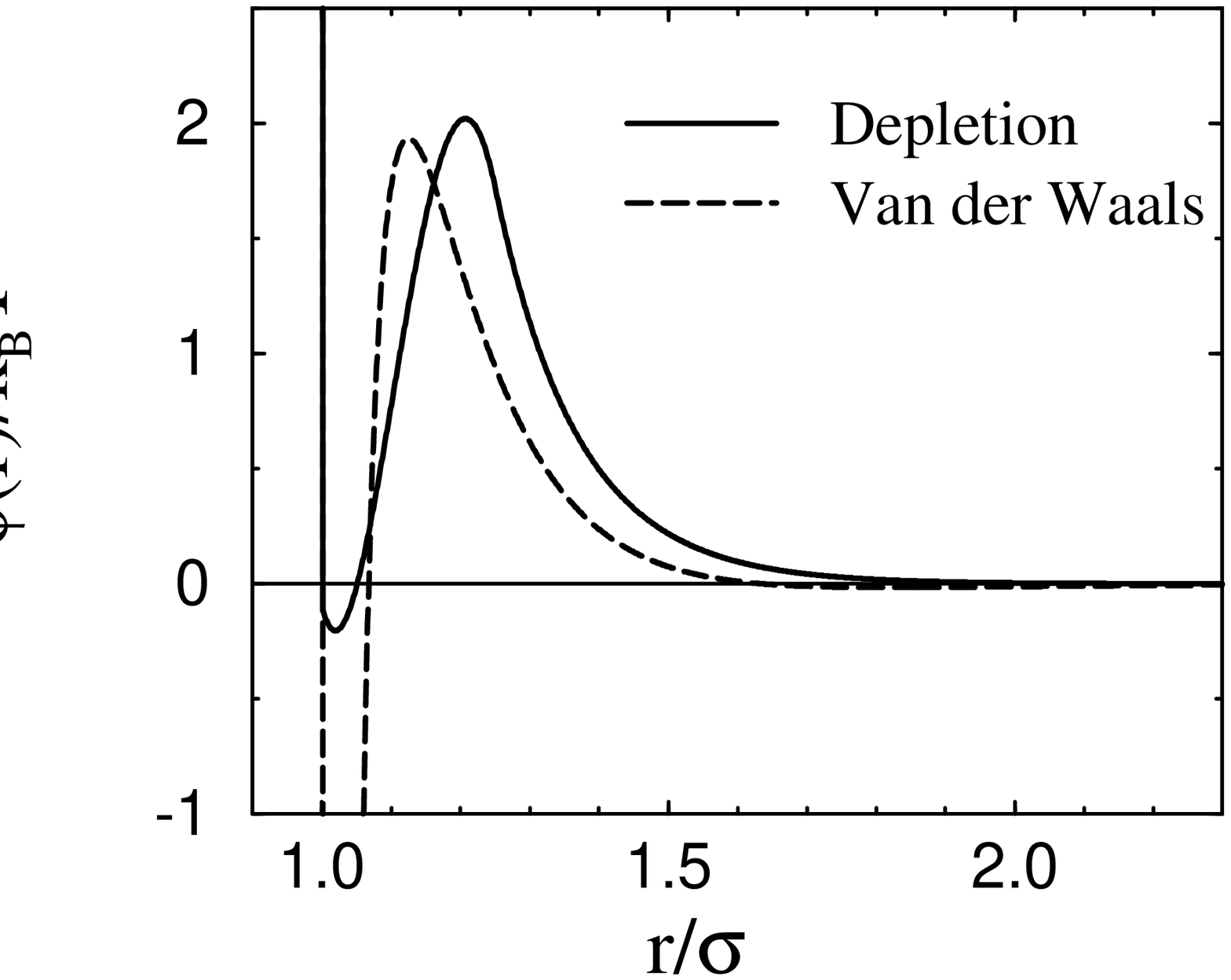,width=140mm,height=100mm}}
\end{picture}
\end{center}
\caption[]{
Effective pair potentials for charge-stabilized colloidal suspensions
with $\sigma=50$ nm, $Z=150 e$.
Solid curve: index-matched suspension mixed with polymer of
$R_g/\sigma=0.125$, $\Pi_p\sigma^3/k_BT=250$.
Dashed curve: non-index-matched with $R_c/\sigma=1.058$, $A/k_BT=25$. 
}
\label{FIG1}
\end{figure}

\begin{figure}
\begin{center}
\begin{picture}(140,100)
\put(5,0){\psfig{figure=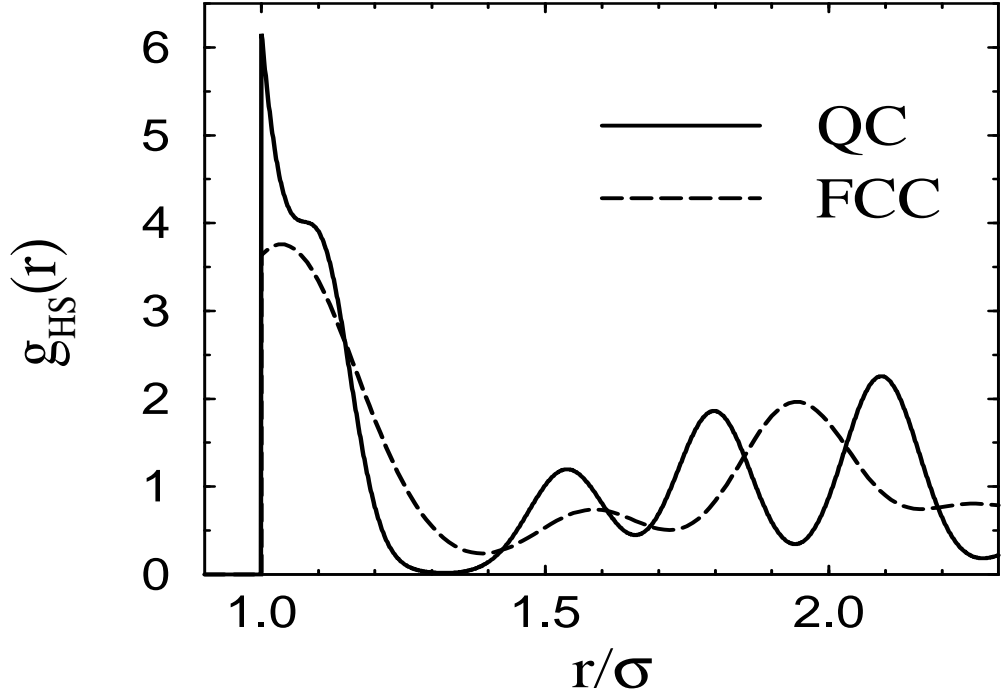,width=140mm,height=100mm}}
\end{picture}
\end{center}
\caption[]{
Radial distribution function of the hard-sphere solid 
for $\eta=0.54$.  
Solid curve:  2/1 rational approximant ($\alpha\sigma^2=230$).
Dashed curve:  FCC crystal ($\alpha\sigma^2=120$). 
}
\label{FIG2}
\end{figure}

\begin{figure}
\begin{center}
\begin{picture}(140,100)
\put(5,0){\psfig{figure=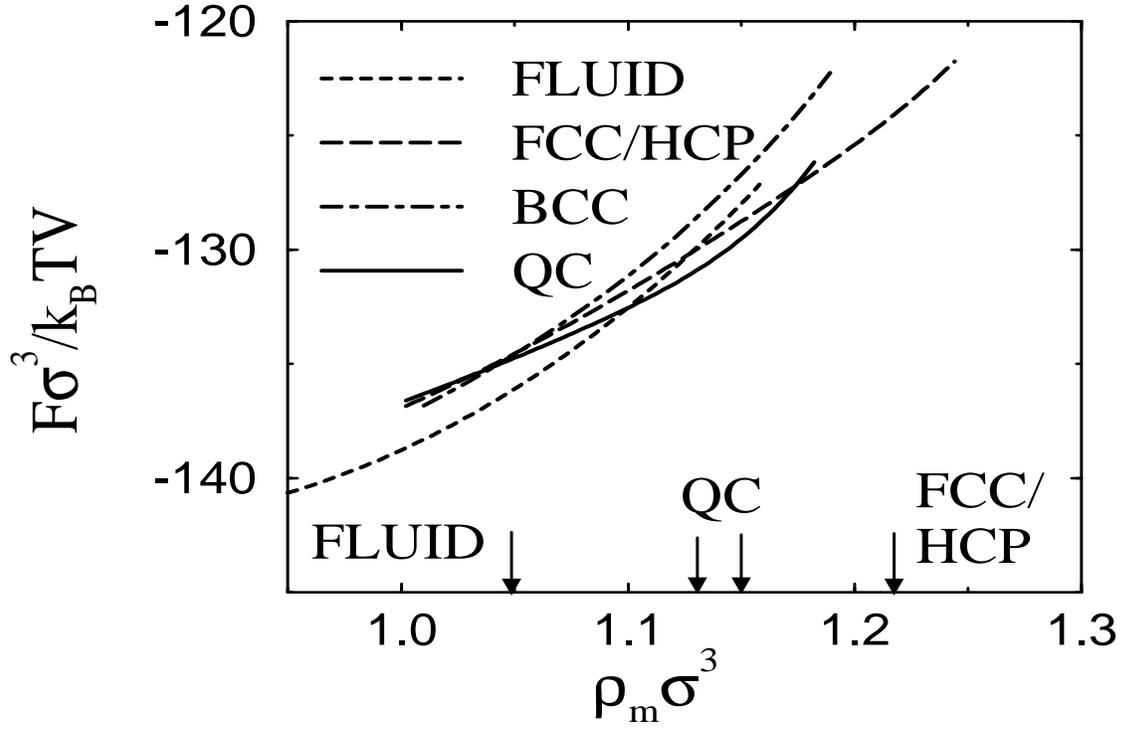,width=140mm,height=100mm}}
\end{picture}
\end{center}
\caption[]{
Helmholtz free energy per unit volume vs. macroion density 
(in reduced units) for a colloidal suspension interacting via 
pair potential of Fig.~1 (dashed curve).
}
\label{FIG3}
\end{figure}

\begin{figure}
\begin{center}
\begin{picture}(140,100)
\put(5,0){\psfig{figure=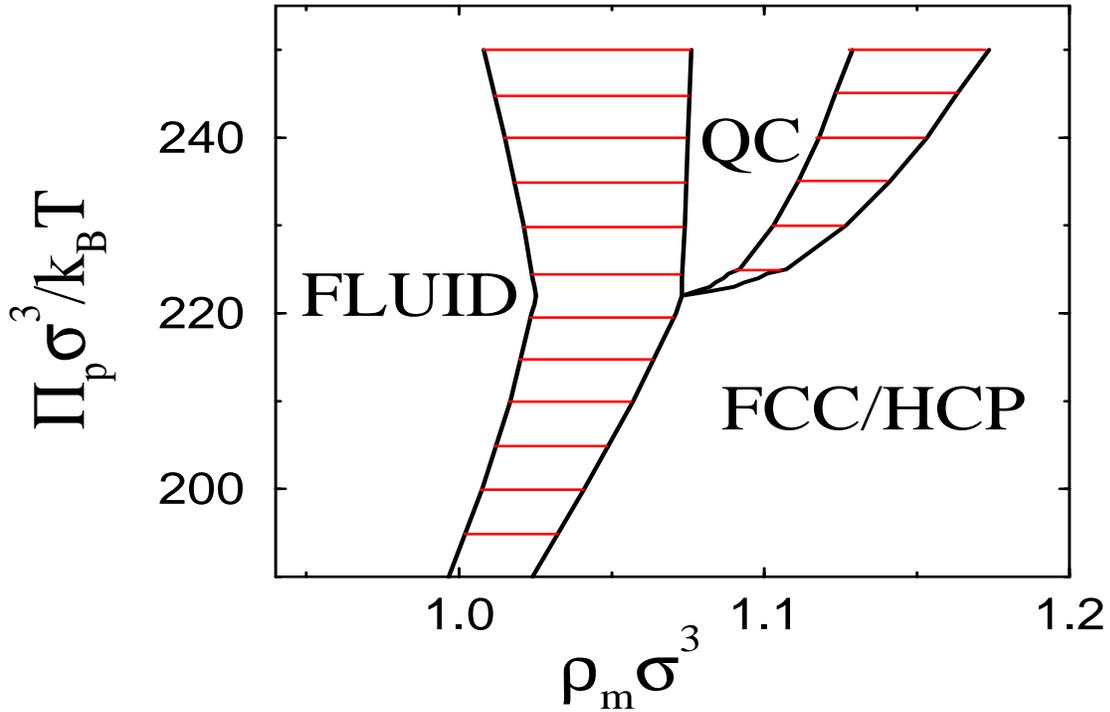,width=140mm,height=100mm}}
\end{picture}
\end{center}
\caption[]{
Phase diagram of polymer osmotic pressure vs. macroion density 
for a colloid-polymer mixture interacting via 
pair potential of Fig.~1 (solid curve).
Horizontal tie lines connect corresponding points on coexistence curves.
}
\label{FIG4}
\end{figure}

\end{document}